\documentclass{article}
\usepackage{spconf,amsmath,graphicx,hyperref}
\usepackage{amssymb}
\usepackage{amsfonts}
\usepackage{booktabs}
\usepackage{subcaption}
\usepackage{multirow}
\usepackage{multicol}
\usepackage{enumitem}
\usepackage{threeparttable}  % 支持表格脚注
\usepackage{booktabs}         % 支持 \toprule, \midrule, \bottomrule

\title{GACA-DiT: Diffusion-based Dance-to-Music Generation with Genre-Adaptive Rhythm and Context-Aware Alignment}

\name{%
  Jinting Wang$^{1}$,
  Chenxing Li$^{2}$,
  Li Liu$^{1}$\sthanks{Corresponding Author: avrillliu@hkust-gz.edu.cn}
}
\address{%
  $^{1}$ The Hong Kong University of Science and Technology (Guangzhou)\\
  $^{2}$ Tencent AI Lab
}

\begin{document}
%\ninept
%
\maketitle
\begin{abstract}
Dance-to-music (D2M) generation aims to automatically compose music that is rhythmically and temporally aligned with dance movements. Existing methods typically rely on coarse rhythm embeddings, such as global motion features or binarized joint-based rhythm values, which discard fine-grained motion cues and result in weak rhythmic alignment. Moreover, temporal mismatches introduced by feature downsampling further hinder precise synchronization between dance and music.
To address these problems, we propose \textbf{GACA-DiT}, a diffusion transformer-based framework with two novel modules for rhythmically consistent and temporally aligned music generation. First, a \textbf{genre-adaptive rhythm extraction} module combines multi-scale temporal wavelet analysis and spatial phase histograms with adaptive joint weighting to capture fine-grained, genre-specific rhythm patterns. Second, a \textbf{context-aware temporal alignment} module resolves temporal mismatches using learnable context queries to align music latents with relevant dance rhythm features. Extensive experiments on the AIST++ and TikTok datasets demonstrate that GACA-DiT outperforms state-of-the-art methods in both objective metrics and human evaluation.
 Project page: \href{ https://beria-moon.github.io/GACA-DiT/}{https://beria-moon.github.io/GACA-DiT/}.

\end{abstract}


% \begin{table}[t]
% \centering
% \caption{Comparison of rhythm modeling strategies in existing D2M methods. 
% Our GACA-DiT introduces a genre-adaptive module that combines multi-scale temporal and spatial analyses with adaptive joint emphasis.}
% \begin{tabular}{p{3cm} p{5cm} p{5.5cm}}
% \toprule
% \textbf{Method} & \textbf{Rhythm Modeling Strategy} & \textbf{Limitations} \\
% \midrule
% D2M-GAN \cite{zhu2022quantized} & Implicit rhythm cues learned from dance–music pairs. & Weak rhythm controllability; prone to misalignment. \\
% CDCD \cite{zhu2022discrete} & Discrete codebook representation of motion–music alignment. & Quantization discards fine-grained rhythm variations. \\
% LORIS \cite{Yu2023Long} & Long-term rhythm regularization with discrete rhythmic states. & Binarized rhythm states hinder precise temporal modeling. \\
% MotionComposer \cite{wang2025motioncomposer} & Motion-conditioned music synthesis with global attention. & Lacks explicit rhythmic feature modeling. \\
% \midrule
% \textbf{GACA-DiT (Ours)} & Multi-scale temporal wavelet analysis + multi-scale spatial phase histograms + adaptive joint weighting. & Preserves fine-grained motion dynamics, ensures genre-aware and robust rhythm embeddings. \\
% \bottomrule
% \end{tabular}
% \label{tab:rhythm_comparison}
% \end{table}

%
\section{Introduction}

Recent advances in multimodal learning have enabled significant progress in generating images, videos, and speech \cite{xiao2025omnigen,henschel2025streamingt2v,10889904}. Among these tasks, music generation has emerged as a key area of interest. In particular, dance-to-music (D2M) generation has gained increasing attention on short video platforms, where it can automatically produce musical accompaniments for user-uploaded dance videos.

\begin{figure}[t]
    \centering
    \includegraphics[width=0.9\linewidth]{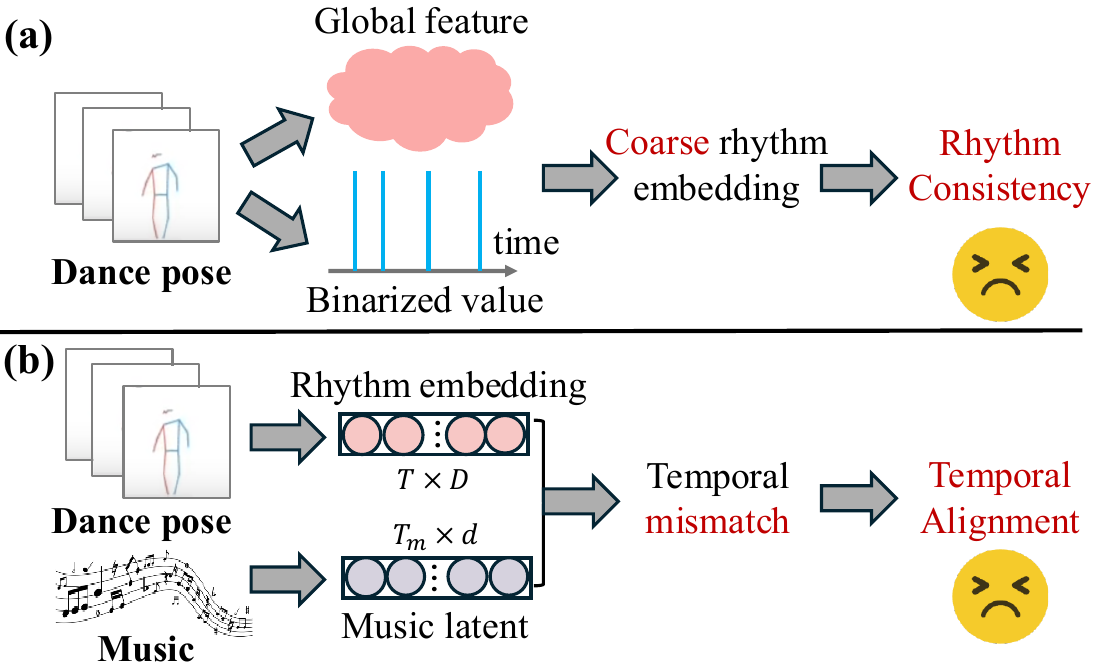}
    \caption{Illustration of key problems in existing D2M methods: coarse rhythm embeddings and temporal misalignment.
}
    \label{fig:motivation}
\end{figure}

D2M aims to generate music that is both temporally synchronized and rhythmically coherent with dance videos. This task is challenging due to two key issues: (1) the difficulty of modeling \textbf{fine-grained rhythmic patterns} across diverse dance genres and (2) the need for precise frame-level \textbf{temporal alignment}.
To tackle the \textbf{first issue}, as illustrated in Figure~\ref{fig:motivation}(a), several studies \cite{zhu2022quantized,liang2024dancecomposer,sun2025enhancing} employ motion encoders such as ST-GCNs \cite{yan2018spatial} to extract global motion features as rhythm representation. While effective at capturing overall motion semantics, these features encode rhythm implicitly and coarsely, thereby limiting rhythmic consistency. In contrast, methods such as LORIS \cite{Yu2023Long} and MotionComposer \cite{wang2025motioncomposer} attempt to explicitly model rhythm by computing joint directions and amplitudes through histograms and discretizing them into binary values. However, such discretization inevitably discards fine-grained rhythmic details that are essential for accurate music-dance alignment.  
As for the \textbf{second issue}, we observe that temporal mismatches between dance rhythm embeddings and music latents, 
caused by downsampling during feature encoding, constitute a critical bottleneck, as visualized in Figure~\ref{fig:motivation}(b). 
As far as we know, this issue has received little attention in prior studies, leaving the alignment between dance and music inadequately addressed.

\begin{figure*}[t]
    \centering
    \includegraphics[width=0.95\linewidth]{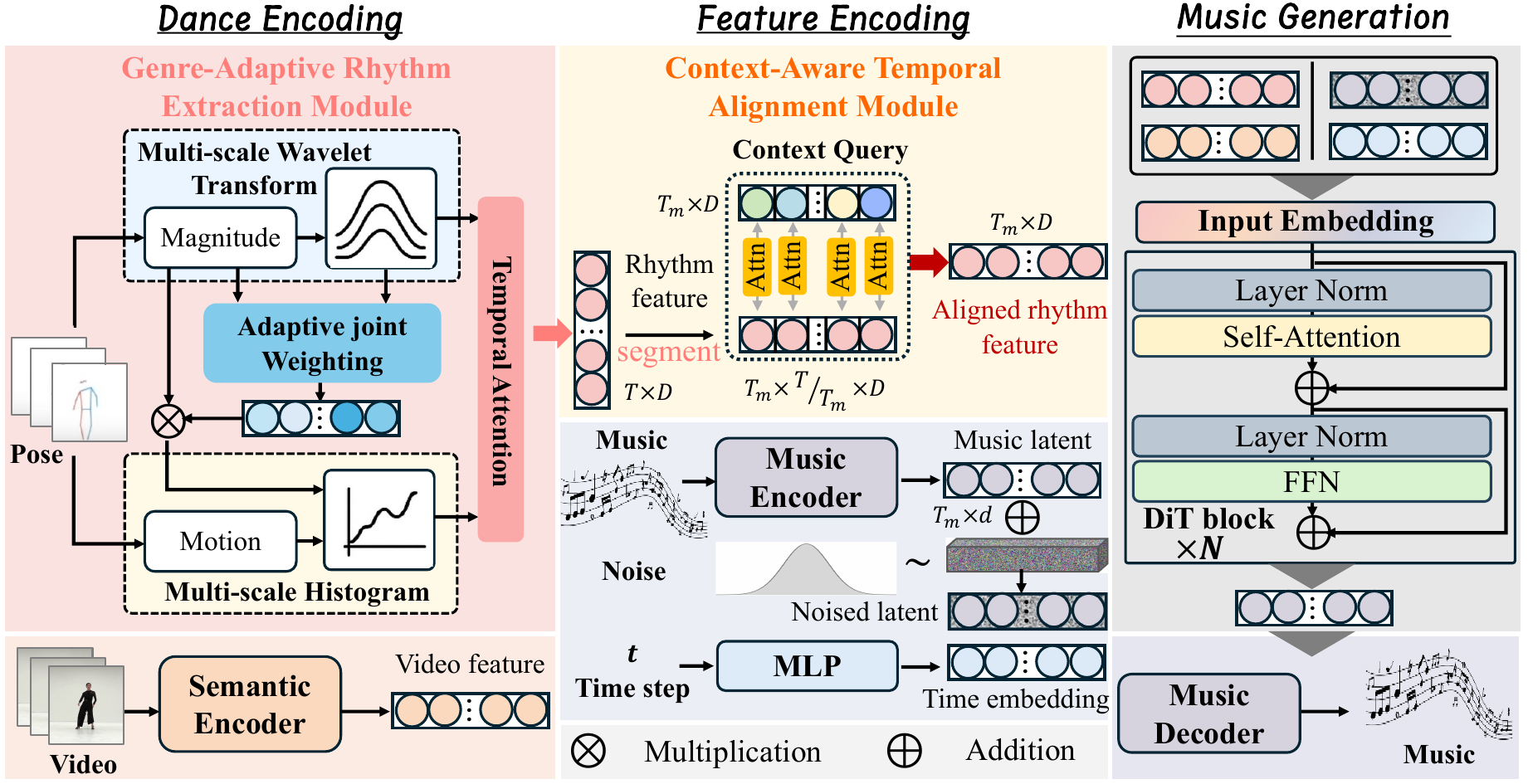}
    \caption{Overview of the proposed \textbf{GACA-DiT} framework. It comprises a genre-adaptive rhythm extraction module, a semantic encoder, a context-aware temporal alignment module, and a Transformer-based diffusion model with music encoder–decoder.}
 
    \label{fig:pipeline}
\end{figure*}

To overcome these limitations, we propose \textbf{GACA-DiT}, a diffusion transformer-based framework with two core innovations for rhythmically consistent and temporally aligned D2M generation. 
\textbf{First}, a \textbf{genre-adaptive rhythm extraction module} jointly leverages multi-scale temporal wavelet analysis and spatial phase histograms as complementary perspectives for rhythm modeling. 
By incorporating adaptive joint weighting and temporal attention, it produces discriminative rhythm embeddings that capture fine-grained motion dynamics across diverse dance genres. 
\textbf{Second}, a \textbf{context-aware temporal alignment module} aligns dense rhythm representations with downsampled music latents using learnable queries guided by contextual information, yielding temporally consistent representations that preserve rhythmic cues for high-fidelity music generation. 

Our main contributions are summarized as follows:
\begin{itemize}
    \item We propose GACA-DiT, a novel diffusion transformer-based framework for D2M generation that enhances rhythmic consistency and temporal alignment.  
    \item We develop a genre-adaptive rhythm extraction module and a context-aware temporal alignment module to capture fine-grained rhythms and resolve cross-modal temporal mismatches.  
    \item Extensive experiments on two dance datasets show that GACA-DiT outperforms state-of-the-art (SOTA) D2M methods in both objective and subjective evaluation.
\end{itemize}

% Extensive experiments on two real-world dance video datasets demonstrate that GACA-DiT outperforms state-of-the-art (SOTA) D2M methods across both objective and subjective metrics.

\section{Related Work of D2M Methods}
\label{sec:related work}

Several approaches have been proposed for D2M generation. Early methods \cite{zhu2022quantized,sun2025enhancing,10539267} embed dance features into music generators using pre-trained motion encoders such as ST-GCN \cite{yan2018spatial}, capturing global motion representations. While effective for overall motion semantics, these methods encode rhythm only implicitly, often resulting in coarse-grained tempo correspondence and weak frame-level synchronization.
Other methods explicitly extract rhythmic cues from pose sequences. For example, RhythmicNet \cite{su2021does}, LORIS \cite{Yu2023Long}, and MotionComposer \cite{wang2025motioncomposer} compute first-order differences of keypoints to obtain motion velocities. These approaches provide more precise rhythm information, but rely on shallow features and short-term dynamics, making them sensitive to noise and less adaptable to diverse dance styles.
To address these limitations, we propose a genre-adaptive rhythm extraction module that captures multi-scale, style-specific motion patterns for robust and fine-grained rhythm modeling.

\section{Method}
\label{sec:method}
\subsection{Overview} 

Figure~\ref{fig:pipeline} illustrates the pipeline of GACA-DiT. 
Given a dance video and corresponding pose sequences, a genre-adaptive rhythm extraction module encodes fine-grained rhythmic patterns into $\mathbf{R} \in \mathbb{R}^{T \times D}$, while an I3D-based semantic encoder extracts video features $\mathbf{V} \in \mathbb{R}^{T \times D_v}$. 
The rhythm features $\mathbf{R}$ are then temporally aligned to the music latent via a context-aware temporal alignment module, producing $\tilde{\mathbf{R}} \in \mathbb{R}^{T_m \times D}$. 
Both $\tilde{\mathbf{R}}$ and $\mathbf{V}$ are then used as conditioning inputs to a diffusion transformer, which generates a music latent that is finally decoded into a waveform rhythmically consistent and semantically aligned with the input dance.

% Figure~\ref{fig:pipeline} presents the pipeline of our proposed GACA-DiT framework. 
% The model first processes the input dance sequence through two parallel encoders: a \textbf{genre-adaptive rhythm extraction (GARE) module}, which produces rhythm features $\mathbf{R} \in \mathbb{R}^{T \times D}$, and an I3D-based semantic encoder, which outputs visual embeddings $\mathbf{V} \in \mathbb{R}^{T \times D_v}$. 
% The rhythm features $\mathbf{R}$ are then temporally aligned to the music latent space via a \textbf{context-aware temporal alignment (CATA) module}, yielding $\tilde{\mathbf{R}} \in \mathbb{R}^{T_m \times D}$. 
% Finally, the aligned rhythm features $\tilde{\mathbf{R}}$ and semantic embeddings $\mathbf{V}$ are used as conditions for a diffusion transformer to generate music that is both rhythmically consistent and semantically aligned with the input dance.

\subsection{Genre-Adaptive Rhythm Extraction}

Accurate rhythm modeling is critical for generating music temporally aligned with dance motion. We propose a \textbf{G}enre-\textbf{A}daptive \textbf{R}hythm \textbf{E}xtraction (\textbf{GARE}) module to explicitly encode fine-grained rhythmic cues from dance pose.
As shown in Figure~\ref{fig:pipeline}, given a dance pose $\mathbf{P} \in \mathbb{R}^{T \times J \times C}$ with $T$ frames, $J$ joints, and $C$ coordinates, we first compute frame-wise motion
$\mathbf{M}_t = \mathbf{P}_{t+1} - \mathbf{P}_t, \quad t=1,\dots,T-1$,
and derive the motion magnitude
$\mathbf{M}^{\text{mag}}_{t,j} = \|\mathbf{M}_{t,j}\|_2$.
For fine-grained dance rhythm, we model dance from complementary \textbf{temporal} and \textbf{spatial} perspectives. We extract multi-scale temporal wavelet features:
\begin{equation}
\mathbf{W}_{t,j,s} = (\mathbf{M}^{\text{mag}}_{:,j} \ast_t \psi_s)_t, \quad s=1,\dots,S,
\end{equation}
where $\ast_t$ denotes 1D convolution along time and $\psi_s$ is the Gabor wavelet kernel at scale $s$. 
To emphasize joints that are more informative for rhythm in each dance genre, we compute joint-adaptive weights:
\begin{equation}
\mathbf{w}_t = \text{softmax}\Big(\text{MLP}([\mathbf{M}^{\text{mag}}_t, \mathbf{W}_t])\Big).
\end{equation}
These weights are then applied to obtain the weighted motion magnitude and the weighted wavelet features. 
The weighted motion magnitude is used to compute multi-scale phase histograms capturing the \textbf{spatial} distribution of motion:
\begin{equation}
\mathbf{h}_{t,k,s} = \sum_j \tilde{\mathbf{M}}^{\text{mag},(s)}_{t,j} \, 
\mathbb{I}\Big[\text{atan2}(M^{y,(s)}_{t,j}, M^{x,(s)}_{t,j}) \in \text{bin}_k\Big],
\end{equation}
where $M^{x,(s)}_{t,j}$, $M^{y,(s)}_{t,j}$ denotes the motion components at scale $s$, $\text{bin}_k$ is the $k$-th phase bin, and $\tilde{\mathbf{M}}^{\text{mag}}_t$ is the weighted motion magnitude.
Finally, histogram features and weighted wavelet features 
are fused via a temporal attention module to produce the rhythm embedding:
\begin{equation}
\mathbf{R}_t = \alpha_t \odot \text{Linear}\Big(\big[\mathbf{h}_t, \sum_j \mathbf{w}_{t,j} \mathbf{W}_{t,j}\big]\Big),
\end{equation}
where $\alpha_t$ is obtained from an attention MLP, $\sum_j \mathbf{w}_{t,j} \mathbf{W}_{t,j}$ is the weighted wavelet features. The final rhythm embedding $\mathbf{R} \in \mathbb{R}^{T \times D}$ is obtained by repeating the last frame, with $D$ embedding dimension.
The rhythm embedding captures fine-grained rhythmic patterns, encoding both multi-scale temporal and spatial dynamics.
% , and provides a conditional signal for music generation.

\subsection{Context-aware Temporal Alignment}

To address the temporal mismatch between the rhythm embedding (temporal length $T$) and the music latent (downsampled length $T_m$), we propose a \textbf{C}ontext-\textbf{A}ware \textbf{T}emporal \textbf{A}lignment (\textbf{CATA}) module.
As shown in Figure~\ref{fig:pipeline}, given rhythm embedding $\mathbf{R} \in \mathbb{R}^{T \times D}$ from the GARE module, we first split it into $T_m$ temporal segments, denoted as $\mathcal{S}_i \in \mathbb{R}^{(T/T_m) \times D}$ for $i = 1, \dots, T_m$. We then employ a set of learnable context queries ~\cite{carion2020end} $\mathbf{Q} = [\mathbf{q}1, \dots, \mathbf{q}{T_m}] \in \mathbb{R}^{T_m\times D}$, where each query $\mathbf{q}_i \in \mathbb{R}^D$ is designed to aggregate contextual information from its corresponding segment $\mathcal{S}_i$ through query-guided attention pooling:
\begin{equation}
\tilde{\mathcal{S}}_i = \sum_j \mathrm{softmax}\Big(\mathbf{r}_j^\top \mathbf{q}_i / \sqrt{D}\Big) \,\mathbf{r}_j,
\end{equation}
where $\mathbf{r}_j$ represents the $j$-th frame feature within segment $\mathcal{S}_i$. This process yields the aligned rhythm representation $\tilde{\mathbf{R}} = [\tilde{\mathcal{S}}_1, \dots, \tilde{\mathcal{S}}_{T_m}] \in \mathbb{R}^{T_m \times D}$, which achieves temporal consistency with the music latents while preserving perceptually important motion dynamics.

\begin{table*}[t]
\centering
\caption{Quantitative comparison on AIST++ and TikTok datasets. The best results are highlighted in bold.}
\label{tab:comparison}
\begin{tabular}{c|c|ccccc|ccccc}
\toprule
\multirow{2}{*}{Method} & \multirow{2}{*}{\shortstack{Params\\(M)}} & \multicolumn{5}{c|}{AIST++} & \multicolumn{5}{c}{TikTok} \\
\cmidrule(lr){3-7} \cmidrule(lr){8-12}
 &  & BCS$\uparrow$ & CSD$\downarrow$ & BHS$\uparrow$ & HSD$\downarrow$ & F1$\uparrow$
 & BCS$\uparrow$ & CSD$\downarrow$ & BHS$\uparrow$ & HSD$\downarrow$ & F1$\uparrow$ \\
\midrule
D2M-GAN \cite{zhu2022quantized} & 63.63  & 89.09 & 14.05 & 88.95 & 22.23 & 88.84 & 82.55 & 28.03 & 87.07 & 24.62 & 84.22 \\
CDCD \cite{zhu2022discrete}     & 428.09 & 92.18 & 14.66 & 80.50 & 21.16 & 84.95 & 84.46 & 25.95 & 87.95 & 23.27 & 86.45 \\
LORIS \cite{Yu2023Long} & 780.16 & 92.13 & 9.94  & 91.71 & 9.69 & 92.05 & 88.02 & 24.14  & 85.25 & 18.72 & 87.43 \\
MotionComposer \cite{wang2025motioncomposer} & 731 & 95.84 & 9.89  & 95.09 & 16.09 & 96.45 & 89.09 & 17.95  & 90.02 & 17.35 & 89.50 \\
GACA-DiT (Ours) & \textbf{56.06} & \textbf{98.13} & \textbf{8.15} & \textbf{98.72} & \textbf{9.93} & \textbf{98.47} 
                 & \textbf{91.55} & \textbf{15.82} & \textbf{91.73} & \textbf{15.56} & \textbf{91.21} \\
\bottomrule
\end{tabular}
\end{table*}

\subsection{Conditional Music Generation}

We formulate D2M generation within a conditional flow matching framework \cite{lipman2023flow}, implemented via a diffusion transformer (DiT). 
The model is conditioned on three inputs: the temporally aligned rhythm feature $\tilde{\mathbf{R}}$, the video features $V$, and the time step $t$. 
It learns a velocity field 
$\mathbf{v}_{\theta}(Z_t, t \mid \tilde{\mathbf{R}}, V)$ that transforms a simple prior $p_0(Z)$ into the target distribution $p_1(Z)$ via the ODE $\frac{d Z_t}{d t} = \mathbf{v}_{\theta}(Z_t, t \mid \tilde{\mathbf{R}}, V)$, with $Z_0 \sim p_0(Z)$ and $Z_1 \sim p_1(Z)$. The model is trained by minimizing the conditional flow matching loss:
\begin{equation}
\mathcal{L}_{\text{CFM}}(\theta) = \mathbb{E}_{t, Z_0, Z_t} 
\Big\|
\mathbf{v}_{\theta}(Z_t, t \mid \tilde{\mathbf{R}}, V) - (Z_1 - Z_0)
\Big\|^2.
\end{equation}

As shown in Figure \ref{fig:pipeline}, during training, ground-truth music waveforms are  first encoded into a latent representation $Z_m \in \mathbb{R}^{T_m \times d}$ using a pre-trained VAE based music encoder. Gaussian noise is added to obtain the noised latent $Z_t$, which is then concatenated with the conditioning features $\{\tilde{\mathbf{R}}, V, t\}$ and fed into the DiT to predict the velocity field.

% During inference, music generation is performed by solving the reverse ODE, starting from a sample $Z_0 \sim p_0(Z)$ and integrating the learned velocity field $\mathbf{v}_{\theta}$ to produce a latent sample $Z_1$ that is both musically coherent and aligned with the input dance. The VAE decoder then synthesizes the final waveform from $Z_1$.

\begin{table}[htbp]
\centering
\begin{threeparttable}
\caption{Ablation study of different components.}
\label{tab:ablation_component}
\begin{tabular}{lccccc}
\toprule
Method & BCS$\uparrow$ & CSD$\downarrow$ & BHS$\uparrow$ & HSD$\downarrow$ & F1$\uparrow$ \\
\midrule
Video Feature & 96.19 & 13.07 & 81.97 & 21.86 & 88.51 \\
+ GARE & 97.52 & 8.19 & 97.71 & 10.02 & 97.51 \\
+ CATA & \textbf{98.13} & \textbf{8.15} & \textbf{98.72} & \textbf{9.93} & \textbf{98.47} \\
\bottomrule
\end{tabular}
% \begin{tablenotes}
% \footnotesize
% \item[*] Progressively added; last row is GACA-DiT.
% \end{tablenotes}
\end{threeparttable}
\end{table}

% \begin{table}[htbp]
% \centering
% \caption{Ablation study of different components.}
% \label{tab:ablation_component}
% \begin{tabular}{lccccc}
% \toprule
% % Method & BCS$\uparrow$ & CSD$\downarrow$ & BHS$\uparrow$ & HSD$\downarrow$ & F1$\uparrow$ \\
% % \midrule
% % Video Feature & 96.19&13.07&81.97&21.86&88.51\\
% % \hline
% %  w/ GARE & 97.52& 8.19& 97.71&10.02&97.51\\
% %  w/ CATA  &\textbf{98.13} & \textbf{8.15} & \textbf{98.72} & \textbf{9.93} & \textbf{98.47}  \\
% % GACA-DiT &\textbf{98.13} & \textbf{8.15} & \textbf{99.13} & \textbf{9.93} & \textbf{98.74}  \\
% \bottomrule
% \end{tabular}
% \end{table}

\begin{table}[htbp]
\centering
\caption{Ablation study of different rhythm features.}
\label{tab:ablation_motion}
\begin{tabular}{lccccc}
\toprule
Method & BCS$\uparrow$ & CSD$\downarrow$ & BHS$\uparrow$ & HSD$\downarrow$ & F1$\uparrow$ \\
\midrule
ST-GCN \cite{yan2018spatial}&96.87& 12.32&95.26&15.82&95.69   \\
prior rhythm \cite{Yu2023Long}& 97.75& 10.43& 96.91&11.08&97.26\\
GARE (Ours) &\textbf{98.13} & \textbf{8.15} & \textbf{98.72} & \textbf{9.93} & \textbf{98.47}  \\
\bottomrule
\end{tabular}
\end{table}

\section{Experiment}
\label{sec:experiment}

\subsection{Experimental Setup}
% \noindent\textbf{Datasets.} 
% We employed two datasets containing paired music and dance videos to evaluate the effectiveness of our GACA-DiT: AIST++ \cite{li2021ai} and TikTok \cite{zhu2022quantized}. 
% We adopt the standard evaluation protocols and dataset splits established by previous baseline methods \cite{zhu2022discrete, sun2025enhancing} to ensure a fair comparison.
\noindent\textbf{Datasets.} 
We evaluate GACA-DiT on two music–dance datasets: AIST++ \cite{li2021ai} and TikTok \cite{zhu2022quantized}, following the standard protocols of prior works \cite{zhu2022discrete, sun2025enhancing} for a fair comparison.

% splits and

% During both training and testing, we adopt the standard evaluation protocols and dataset splits established by previous baseline methods \cite{zhu2022discrete, sun2025enhancing} to ensure a fair comparison.

\noindent\textbf{Implementation Details.} 
We employ a pre-trained I3D model \cite{yadav2022inflated} (trained on ImageNet \cite{deng2009imagenet}) as the semantic encoder to extract video features from video frames, and a pre-trained VAE from DiffRhythm \cite{ning2025diffrhythm} to encode and reconstruct audio waveforms. The conditional DiT comprises 8 transformer blocks, each with a 512-dimensional hidden size and 10-head self-attention. Experiments use 5-second music clips at 44,100 Hz, and 2D pose skeletons are obtained via DWpose \cite{yang2023effective}. We train with a batch size of 4 for 100 epochs using the Adam optimizer ($\beta_1=0.9$, $\beta_2=0.95$) with a learning rate of $1\times10^{-4}$. The diffusion process uses a 32-step Euler ODE solver, and classifier-free guidance \cite{ho2021classifier} with a scale of 4 is applied during inference.

\noindent\textbf{Evaluation Metrics.}  
For \textbf{objective evaluation}, following \cite{Yu2023Long}, we measure the alignment between musical rhythms and dance patterns using Beats Coverage Score (BCS), Beats Hit Score (BHS), their F1 scores, and the standard deviations of BCS and BHS, denoted as CSD and HSD, respectively.  
For \textbf{subjective evaluation}, we use the Mean Opinion Score (MOS) to assess both the overall quality of the generated music and its rhythm consistency with the dance video, on a scale from 1 (poor) to 5 (excellent).

\subsection{Compared with SOTAs} 
\begin{figure}[htbp]
    \centering
    \includegraphics[width=0.9\linewidth]{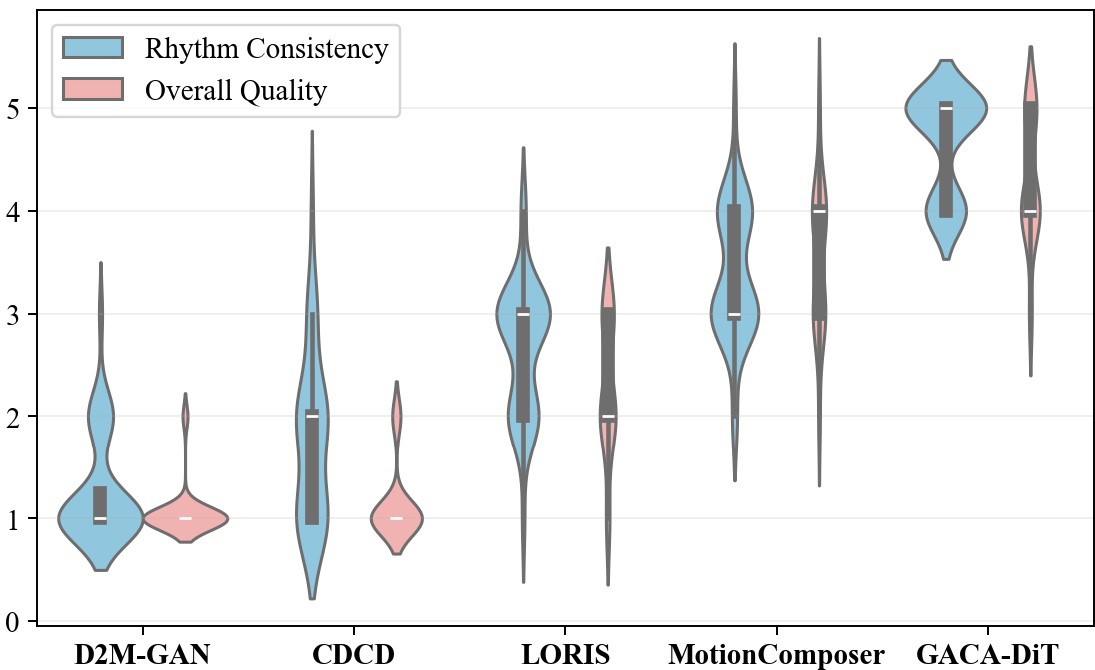}
    \caption{Violin plots of MOS results for subjective evaluation. }
    \label{fig:user_study}
\end{figure}

To validate the effectiveness of our proposed GACA-DiT, we compare it with several D2M methods, including D2M-GAN \cite{zhu2022quantized}, CDCD \cite{zhu2022discrete}, LORIS \cite{Yu2023Long}, and MotionComposer \cite{wang2025motioncomposer}. The quantitative results are summarized in Table \ref{tab:comparison}, showing that GACA-DiT consistently outperforms all baseline methods across all metrics on both datasets. 
In addition, we conduct a user study for subjective evaluation, inviting 20 volunteers to rate 30 test samples from the AIST++ dataset generated by the five D2M methods. As shown in the violin plots in Figure \ref{fig:user_study}, GACA-DiT achieves higher median scores and more concentrated distributions for both rhythmic consistency and overall quality, indicating that its generated results are not only well-aligned with the input dance but also consistently of high audio quality.

\subsection{Ablation Analysis}
\noindent\textbf{Model Components.} 
We conduct ablation studies to evaluate the contributions of the proposed components in GACA-DiT, as listed in Table~\ref{tab:ablation_component}. Compared with using only video features, using the fine-grained rhythm embedding from the GARE module significantly improves all metrics, demonstrating the key role of rhythmic cues in aligning dance with music. Further incorporating the CATA module enhances temporal alignment, resulting in more coherent music generation.

\noindent\textbf{Rhythm Feature.} 
Table~\ref{tab:ablation_motion} compares the impact of different rhythm features on GACA-DiT. The proposed GARE module outperforms both ST-GCN \cite{yan2018spatial} and the rhythm extraction method used in LORIS \cite{Yu2023Long}, demonstrating its superior ability to capture fine-grained rhythmic cues and achieve more accurate dance-to-music alignment.

\vspace{-4mm}

\section{Conclusion}
\label{sec:conclusion}

In this paper, we introduced GACA-DiT, a diffusion-based framework for dance-to-music generation that improves both rhythmic consistency and temporal alignment. Our approach integrates a genre-adaptive rhythm extraction module to capture fine-grained rhythmic patterns and a context-aware temporal alignment module to mitigate cross-modal temporal mismatches. Extensive experiments on the AIST++ and TikTok datasets show that GACA-DiT consistently surpasses SOTAs, achieving superior performance in both objective metrics and perceptual quality.
% in terms of both objective metrics and perceptual quality.

\newpage
\bibliographystyle{IEEEbib}
\bibliography{strings,refs}

\end{document}